# Quantum Boolean Image Denoising

**Mario Mastriani**

DLQS LLC, 4431 NW 63RD Drive, Coconut Creek, FL 33073, USA.
mmastri@gmail.com

*Abstract*—A quantum Boolean image processing methodology is presented in this work, with special emphasis in image denoising. A new approach for internal image representation is outlined together with two new interfaces: classical-to-quantum and quantum-to-classical. The new quantum-Boolean image denoising called quantum Boolean mean filter (QBMF) works with computational basis states (CBS), exclusively. To achieve this, we first decompose the image into its three color components, i.e., red, green and blue. Then, we get the bitplanes for each color, e.g., 8 bits-per-pixel, i.e., 8 bitplanes-per-color. From now on, we will work with the bitplane corresponding to the most significant bit (MSB) of each color, exclusive manner. After a classical-to-quantum interface (which includes a classical inverter), we have a quantum Boolean version of the image within the quantum machine. This methodology allows us to avoid the problem of quantum measurement, which alters the results of the measured except in the case of CBS. Said so far is extended to quantum algorithms outside image processing too. After filtering of the inverted version of MSB (inside quantum machine) the result passes through a quantum-classical interface (which involves another classical inverter) and then proceeds to reassemble each color component and finally the ending filtered image. Finally, we discuss the more appropriate metrics for image denoising in a set of experimental results.

*Keywords*—Quantum algorithms - Quantum-Boolean Image Denoising - Quantum/Classical Interfaces - Quantum measurement.

## 1 Introduction

Quantum computation and quantum information is the study of the information processing tasks that can be accomplished using quantum mechanical systems. Like many simple but profound ideas it was a long time before anybody thought of doing information processing using quantum mechanical systems [1].

Quantum computation is the field that investigates the computational power and other properties of computers based on quantum-mechanical principles. An important objective is to find quantum algorithms that are significantly faster than any classical algorithm solving the same problem. The field started in the early 1980s with suggestions for analog quantum computers by Paul Benioff [2] and Richard Feynman [3, 4], and reached more digital ground when in 1985 David Deutsch defined the universal quantum Turing machine [5]. The following years saw only sparse activity, notably the development of the first algorithms by Deutsch and Jozsa [6] and by Simon [7], and the development of quantum complexity theory by Bernstein and Vazirani [8]. However, interest in the field increased tremendously after Peter Shor's very surprising discovery of efficient quantum algorithms (or simulations on a quantum computer) for the problems of integer factorization and discrete logarithms in 1994 [9].

Since most of current classical cryptography is based on the assumption that these two problems are computationally hard, the ability to actually build and use a quantum computer would allow us to break most current classical cryptographic systems, notably the Rivest, Shamir y Adleman (RSA) system [10, 11]. In contrast, a quantum form of cryptography due to Bennett and Brassard [12] is unbreakable even for quantum computers.

On the other hand, and as well say Hirota *et al* inside the Introduction of their work [13]:

*Quantum computation has appeared in various areas of computer science such as information theory, cryptography, image processing, etc. [1] because there are inefficient tasks on classical computers that can*

*be overcome by exploiting the power of the quantum computation. Processing and analysis of images in particular and visual information in general on classical computers have been studied extensively [14-17]. On quantum computers, the research on images has faced fundamental difficulties because the field is still in its infancy. To start with, what are quantum images or how do we represent images on quantum computers? Secondly, what should we do to prepare and process the quantum images on quantum computers?*

Precisely, these two questions represent the essence on which this paper is based, i.e., the correct (and more efficient) internal representation of an image in a quantum context, and its recovery, once processed internally. Thus, we recognize only 3 milestones in the brief history of quantum image processing, namely:

- all starts with the pioneering work of Prof. Salvador E. Venegas-Andraca [18-21] at Keble College, Oxford University, UK (currently at Tecnológico de Monterrey, Campus Estado de México), where he proposes quantum image representations such as Qubit Lattice [22], in fact, this is the first doctoral thesis in the specialty,

- the history continues with the quantum image representation via the Real Ket [23] of Prof. Jose I. Latorre Sentís, at Universitat de Barcelona, Spain, with a special interest in image compression in a quantum context, and finally,

- we arrive at the proposal of Prof. Kaoru Hirota *et al* [13] from Tokyo Institute of Technology, for a flexible representation of quantum images to provide a representation for images on quantum computers in the form of a normalized state which captures information about colors and their corresponding positions in the images.

These works marked the path and viability of quantum image processing, however, we believe that a new type of internal representation of images, which enable an easier representation of traditional algorithms of traditional Digital Image Processing in a quantum computer, as well as more easy and efficient recovery of images processed outside the quantum computer is imperative. This is the essence of this work, which is organized as follows:

The basic principles of Quantum Information Processing are outlined in Section 2. Implementation Problems in Quantum Image Processing are presented in Section 3. The new approach for internal image representation is outlined in Section 5, where, we present the development of Quantum-Boolean Image Processing concept. Besides, in this section, we show the proposed new interfaces classical-to-quantum and quantum-to-classical, and a new quantum-Boolean image denoising called quantum Boolean mean filter (QBMF). In Section 5, we discuss the more appropriate metrics for image denoising in a set of experimental results. Finally, Section 6 provides a conclusion and future works proposal of the paper.

## 2 Quantum Information Processing

In this section, we present the main concepts related to Quantum Information Processing, that is to say: qubit, Bloch's Sphere, Hilbert's Space, Schrödinger Equation, what happens before and after Quantum Measurement, Unitary Operators, Quantum Circuits/Gates, and Quantum Algorithms.

2.1 Quantum bits (qubits) and Bloch's sphere

The bit is the fundamental concept of classical computation and classical information. Quantum computation and quantum information are built upon an analogous concept, the quantum bit, or qubit for short. In this section we introduce the properties of single and multiple qubits, comparing and contrasting their properties to those of classical bits [1].

The difference between bits and qubits is that a qubit can be in a state other than $|0\rangle$ or $|1\rangle$ [24, 25]. It is also possible to form linear combinations of states, often called superpositions:

$$|\psi\rangle = \alpha|0\rangle + \beta|1\rangle, \tag{1}$$

where $|\alpha|^2 + |\beta|^2 = 1$, with the states $|\alpha\rangle$ and $|\beta\rangle$ are understood as different polarization states of light. The numbers α and β are complex numbers, although for many purposes not much is lost by thinking of them as real numbers. Put another way, the state of a qubit is a vector in a two-dimensional complex vector space. The special states $|0\rangle$ and $|1\rangle$ are known as Computational Basis States (CBS), and form an orthonormal basis for this vector space, being $|0\rangle = \begin{bmatrix} 1 \\ 0 \end{bmatrix}$ and $|1\rangle = \begin{bmatrix} 0 \\ 1 \end{bmatrix}$

One picture useful in thinking about qubits is the following geometric representation.

Because $|\alpha|^2 + |\beta|^2 = 1$, we may rewrite Equation (1) as

$$|\psi\rangle = e^{i\gamma}\left(\cos\frac{\theta}{2}|0\rangle + e^{i\phi}\sin\frac{\theta}{2}|1\rangle\right) = e^{i\gamma}\left(\cos\frac{\theta}{2}|0\rangle + (\cos\phi + i\sin\phi)\sin\frac{\theta}{2}|1\rangle\right) \tag{2}$$

where $0 \leq \theta \leq \pi$, $0 \leq \phi < 2\pi$. We can ignore the factor of $e^{i\gamma}$ out the front, because it has no observable effects [1], and for that reason we can effectively write

$$|\psi\rangle = \cos\frac{\theta}{2}|0\rangle + e^{i\phi}\sin\frac{\theta}{2}|1\rangle \tag{3}$$

The numbers θ and φ define a point on the unit three-dimensional sphere, as shown in Fig. 1.

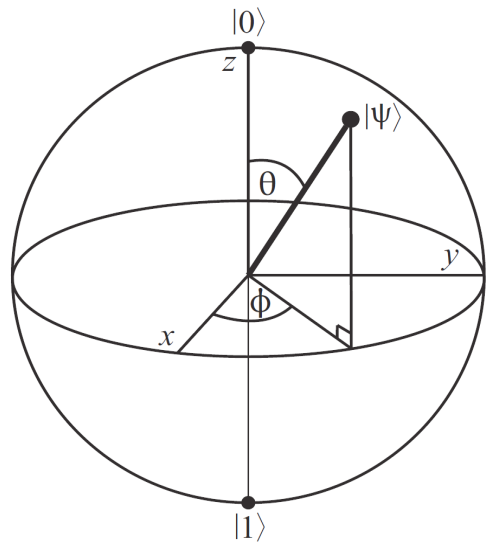

**Fig. 1** Bloch's Sphere.

2.2 Schrödinger´s equation and quantum algorithms

Given the following differential equation known as the Schrödinger equation time dependent [1, 24-26]:

$$\frac{d}{dt}|\psi(t)\rangle = \frac{-i\hat{H}}{\hbar}|\psi(t)\rangle \tag{4}$$

where $\hat{H}$ represents the Hamiltonian matrix of the Schrödinger equation, $i = \sqrt[2]{-1}$, and $\hbar$ is the Planck constant, being $|\psi(t)\rangle$ the wave function, thus the probability amplitudes evolve across time according to the following equation:

$$|\psi(t)\rangle = e^{\frac{-i\hat{H}t}{\hbar}}|\psi(0)\rangle \quad (5)$$

Equation 5 is the main piece in building circuits, gates and quantum algorithms [1]. On the other hand, If we make $|\psi(t)\rangle = U(t)|\psi(0)\rangle$, being $U(t) = e^{\frac{-i\hat{H}t}{\hbar}}$, with $U^{\dagger}U = I$ (where $I$ is the identity matrix), in particular, unitary operators will preserve lengths: $\langle\psi|U^{\dagger}U|\psi\rangle = \langle\psi|\psi\rangle = 1$, this is very important.

2.3 Quantum Circuits, Gates and Algorithms

As we can see in Fig. 2, and remember Eq.(5), the quantum algorithm (identical case to circuits and gates) viewed as a transfer (or mapping input-to-output) has two types on output:

*a)* the result of algorithm (circuit of gate), i.e., $|\psi_{out}\rangle$

*b)* part of the input $|\psi_{in}\rangle$, i.e., $|\underline{\psi}_{in}\rangle$ (underlined $|\psi_{in}\rangle$), in order to impart reversibility to the circuit, which is a critical need in quantum computing [1].

Besides, we can see clearly a module for measuring $|\psi_{out}\rangle$ (which will be extensively discussed in the next section) with their respective output, i.e., $|\psi_{out}\rangle_{pm}$ (where subscript *pm* means post-measurement), and a number of elements needed for the physical implementation of the quantum algorithm (circuit or gate), namely: control, ancilla and trash [1]. In this figure as well as in the rest of them (unlike [1]) a single fine line represents a wire carrying *1* qubit or *N* qubits (qudit), interchangeably, while a single thick line represents a wire carrying *1* or *N* classical bits, interchangeably too.

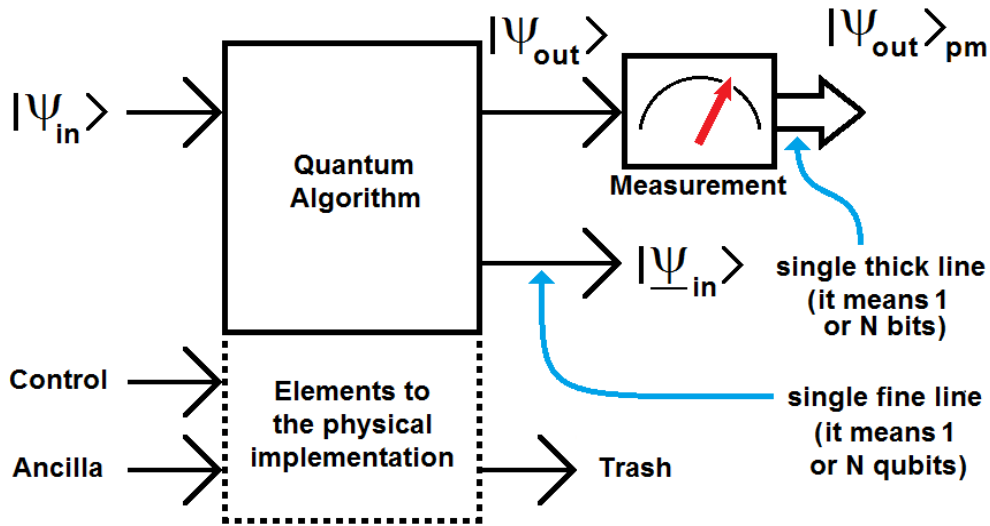

**Fig. 2** Module to measuring, quantum algorithm and the elements needs to its physical implementation.

However, the mentioned concept of reversibility is closely related to energy consumption, and hence to the Landauer's Principle [1]. On the other hand, computational complexity studies the amount of time and space required to solve a computational problem. Another important computational resource is energy. In [1], the authors show the energy requirements for computation. Surprisingly, it turns out that computation, both classical and quantum, can in principle be done without expending any energy! Energy consumption in computation turns out to be deeply linked to the reversibility of the computation. In other words, it is inexcusable the need of the $|\underline{\psi}_{in}\rangle$ presence to the output of quantum gate [1].

# 3 Implementation Problems in Quantum Image Processing

The implementation problems in Quantum Image Processing are: Wave function collapse, Quantum Measurement Problems (pre-measurement and post-measurement quantum states are different in general [22]), Types of measurement and state reconstruction, Interfaces, and Internal representations of an image and its possible implementations. However, we present here the most important of them relatives to carry out quantum logic operations with CBS, which are fundamental concepts for the posterior development of our own internal representation of an image (inside quantum processor), classical-to-quantum and quantum-to-classical interfaces, i.e., the difference between wave function before and after quantum measurement.

***Postulate.*** Quantum measurements are described by a set of measurement operators $\{\hat{M}_m\}$, index $m$ labels the different measurement outcomes, which act on the state space of the system being measured. Measurement outcomes correspond to values of *observables*, such as position, energy and momentum, which are Hermitian operators [1, 22] corresponding to physically measurable quantities.

Let $|\psi\rangle$ be the state of the quantum system immediately before the measurement. Then, the probability that result $m$ occurs is given by

$$p(m) = \langle \psi | \hat{M}_m^\dagger \hat{M}_m | \psi \rangle \qquad (6)$$

and the post-measurement quantum state is

$$|\psi\rangle_{pm} = \frac{\hat{M}_m |\psi\rangle}{\sqrt{\langle \psi | \hat{M}_m^\dagger \hat{M}_m | \psi \rangle}} \qquad (7)$$

Operators $\hat{M}_m$ must satisfy the completeness relation of Eq.(8a), because that guarantees that probabilities will sum to one; see Eq.(8b) [22]:

$$\sum_m \hat{M}_m^\dagger \hat{M}_m = I \qquad (8a)$$
$$\sum_m \langle \psi | \hat{M}_m^\dagger \hat{M}_m | \psi \rangle = \sum_m p(m) = 1 \qquad (8b)$$

Let us work out a simple example. Assume we have a polarized photon with associated polarization orientations 'horizontal' and 'vertical'. The horizontal polarization direction is denoted by $|0\rangle$ and the vertical polarization direction is denoted by $|1\rangle$. Thus, an arbitrary initial state for our photon can be described by the quantum state $|\psi\rangle = \alpha|0\rangle + \beta|1\rangle$, where $\alpha$ and $\beta$ are complex numbers constrained by the normalization condition $|\alpha|^2 + |\beta|^2 = 1$ and $\{|0\rangle, |1\rangle\}$ is the computational basis spanning $H^2$. Now, we construct two measurement operators $\hat{M}_0 = |0\rangle\langle 0|$ and $\hat{M}_1 = |1\rangle\langle 1|$ and two measurement outcomes $a_0$, $a_1$. Then, the full observable used for measurement in this experiment is $\hat{M} = a_0|0\rangle\langle 0| + a_1|1\rangle\langle 1|$. According to Postulate, the probabilities of obtaining outcome $a_0$ or outcome $a_1$ are given by $p(a_0) = |\alpha|^2$ and $p(a_1) = |\beta|^2$. Corresponding post-measurement quantum states are as follows: if outcome = $a_0$ then $|\psi\rangle_{pm} = |0\rangle$; if outcome = $a_1$ then $|\psi\rangle_{pm} = |1\rangle$. Finally, in quantum mechanics, measurement is a non-trivial and highly counter-intuitive process. Firstly, because measurement outcomes are inherently probabilistic, i.e. regardless of the carefulness in the preparation of a measurement procedure, the possible outcomes of such measurement will be distributed according to a certain probability distribution. Secondly, once the measurement has been performed, a quantum system in unavoidably altered due to the interaction with the measurement apparatus, i.e., after measuring the wave function collapses hopelessly except for CBS.

# 4  Quantum-Boolean Image Processing (QuBoIP)

QuBoIP is presented as a branch of Quantum Image Processing that is composed of the following steps, namely:

- *color decomposition and bit slicing*
- *classical-to-quantum interface (C2QI)*
- *quantum Boolean image denoising*
- *quantum-to-classical interface (Q2CI)*
- *bit reassembling and color recomposition*

## 4.1  Color decomposition and bit slicing

We decompose the original noisy image in its color components (i.e., red, green and blue), and in turn, each color component in their corresponding bitplanes thanks to bit slicing, in this case thanks to an own MATLAB® function [27] called *slicer(.)*, and from which we get many bitplanes as depth in bit has the image to be treated. In Fig.3, we get 8 bitplanes, where, bitplane 7 is called Most Significant Bit (MSB) and it is the most morphologically committed bitplane with the original image [28]. In return, bitplane 0 is the Least Significant Bit (LSB) and it is the least morphologically committed bitplane with the gray image.

Two important aspects:

- From here to the end of this paper, we are going to work with MSB, i.e., with we will say "image", we are saying MSB.

- The classical version of the *slicer(.)* function in MATLAB® code is:

```
function Ibpp = slicer(I,bpp)                    function bvpp = d2b(p,bpp)

% Casting of algorithm:                          % Casting of algorithm:
% bpp = bit-per-pixel                            % d = bit depth
% I = Each color component of the image          % p = pixel value
% Ibpp = I in bpp bitplanes (strictly binary)    % bvpp = binary vector per pixel

[ROW,COL] = size(I);                             bvpp = zeros(1,bpp);
for r = 1:ROW                                    d = 1;
  for c = 1:COL                                  while p > 0,
    aux = d2b(I(r,c)-1,bpp);                       bvpp(d) = mod(p,2);
    for b = 1:bpp                                  p = p/2;
      Ibpp(r,c,b) = aux(b);                        p = floor(p);
    end                                            d = d+1;
  end                                            end
end                                              bvpp = rot90(rot90(bvpp));

return;                                          return;
```

If we were to highlight the advantage of working in QuBoIP rather than in QuIP [13, 18-23], it would certainly be the fact that as QuBoIP working with CBS exclusively, the measurement is not a problem as in the rest of Quantum Physics, because, when we measured an $|1\rangle$, the result is an unchanged $|1\rangle$, and when we measured an $|0\rangle$, the result is an unchanged $|0\rangle$.

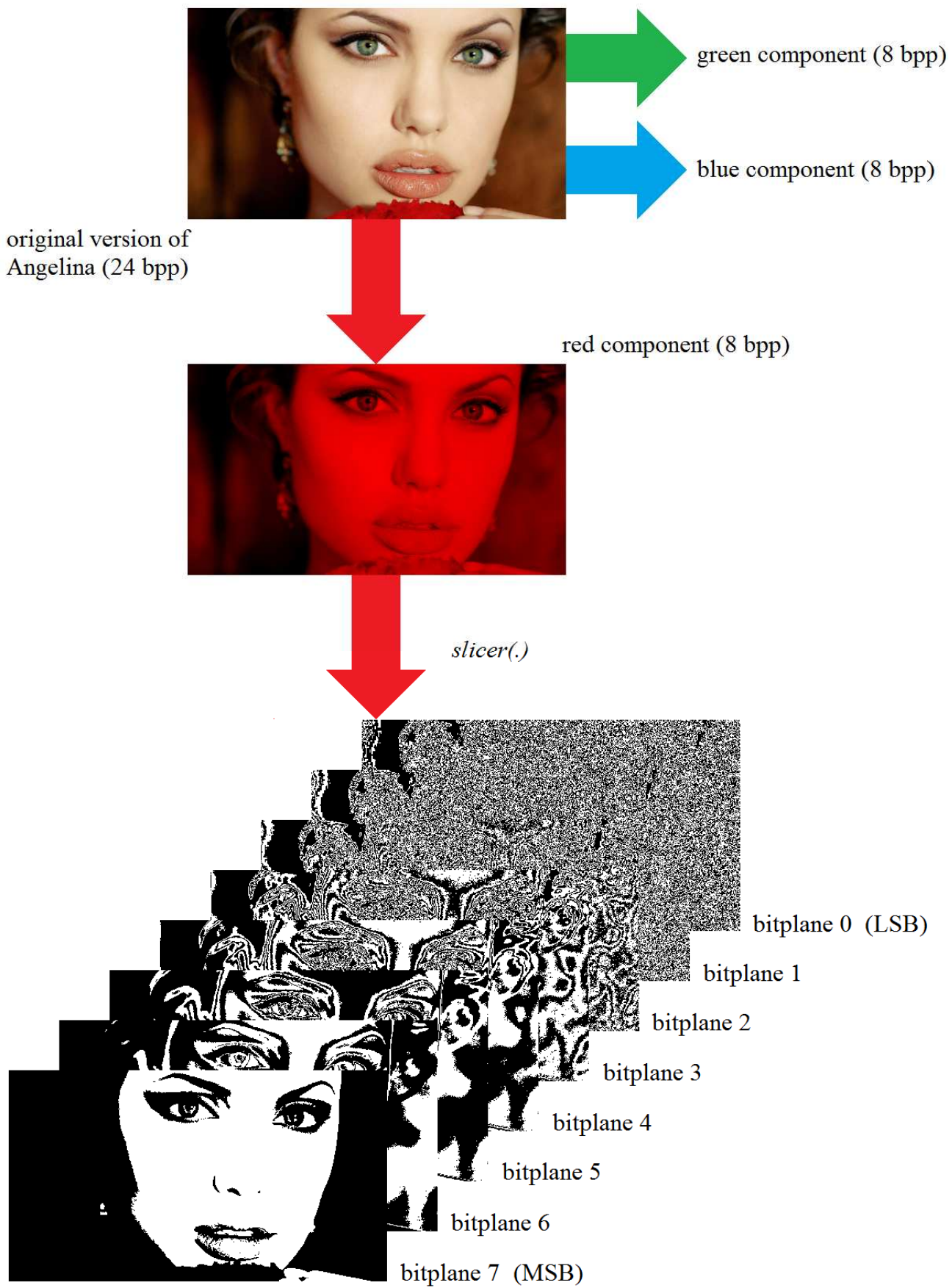

**Fig. 3** Bitplanes of the red component for Angelina obtained by slicing, with special remarks for MSB and LSB.

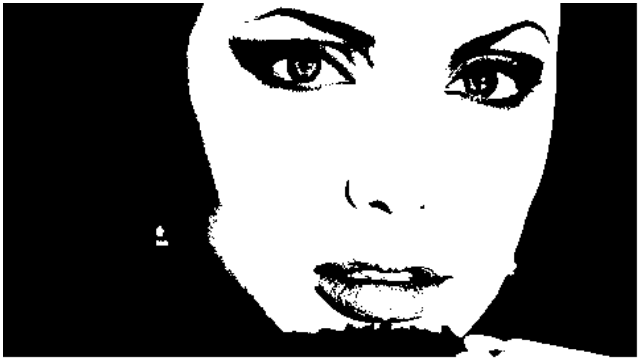
(MSB) bitplane 7

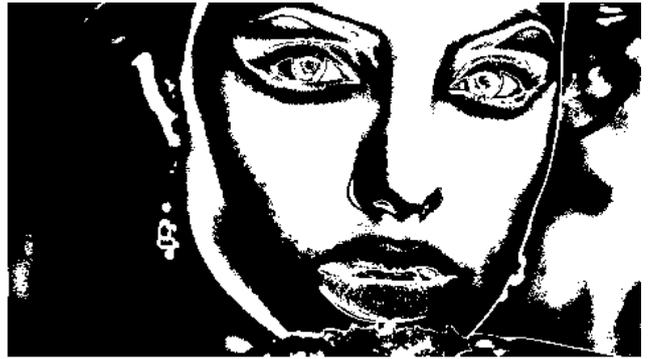
bitplane 6

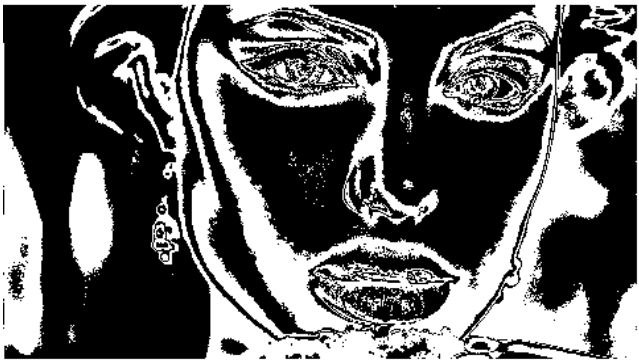
bitplane 5

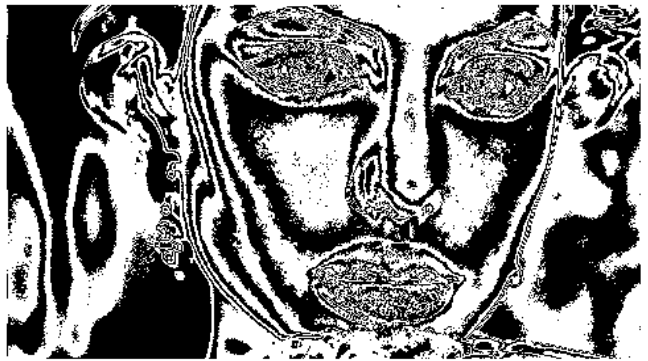
bitplane 4

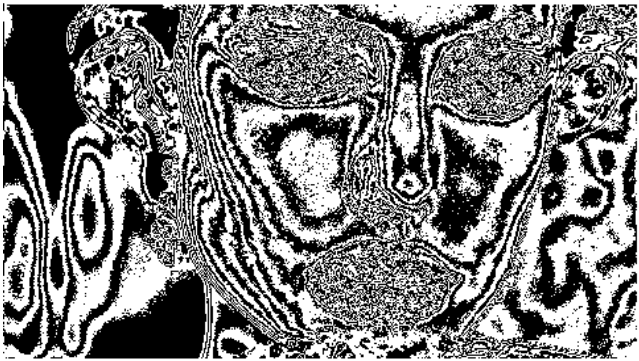
bitplane 3

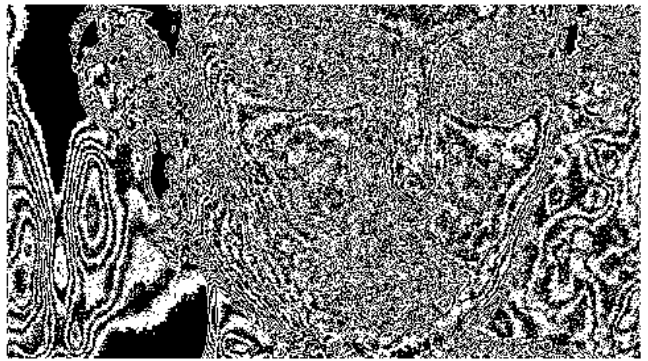
bitplane 2

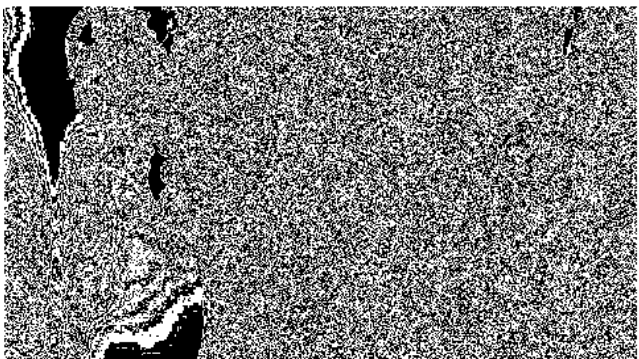
bitplane 1

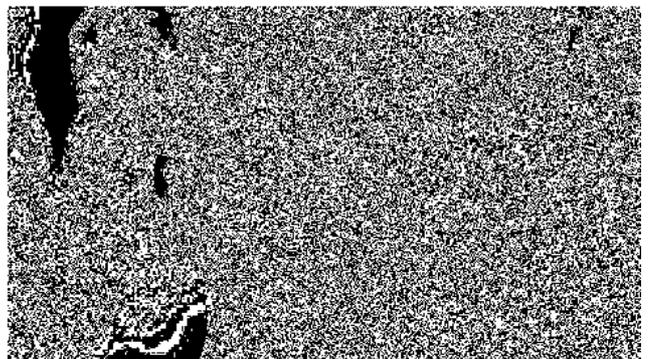
(LSB) bitplane 0

**Fig. 4** Angelina and her 8 bitplanes, including MSB and LSB.

Figure 4 show us -in detail- the 8 bitplanes of Angelina, from MSB (bitplane 7) to LSB (bitplane 0). Let observe that as we move from MSB to LSB, different bitplanes are increasingly unrecognizable compared to the original image, i.e., Angelina. As we can see, LSB is completely different regarding to original Angelina morphology. This is one reason why the LSB is Steganography territory [28]. The other reason is that any change in the LSB does not produce visually detectable changes in the original image.

### 4.2 Classical-to-quantum interface (C2QI)

In this section, a complete description of the operating principle of this interface is presented. This includes the relationship between external and internal representation of MSB (bitplane 7) for each color.

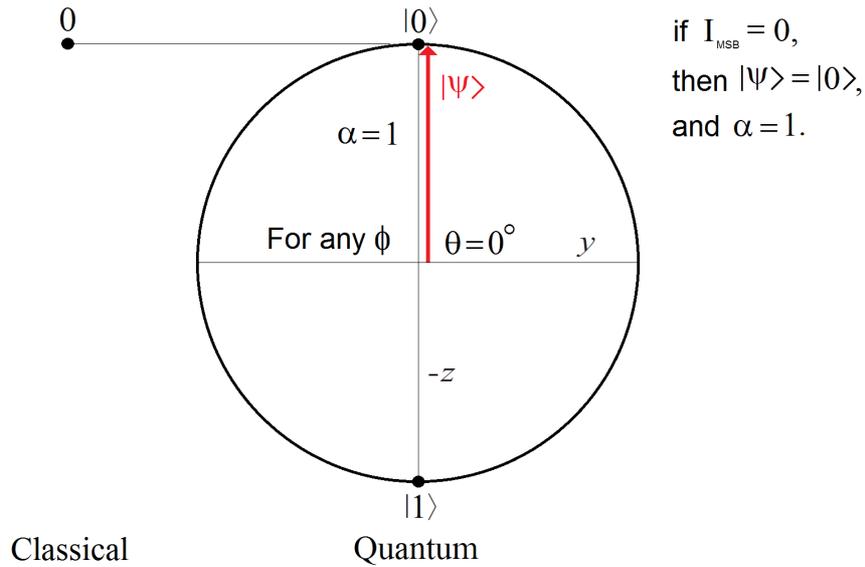

**Fig. 5** Relationship between classical 0, $\alpha$ and $I_{MSB}$.

Figure 5 show as the relationship between classical 0, $\alpha$ and $I_{MSB}$. In this case, $\alpha = 1$ when $I_{MSB} = 0$, and $|\psi\rangle = |0\rangle$, with $\theta = 0°$ and for any $\phi$.

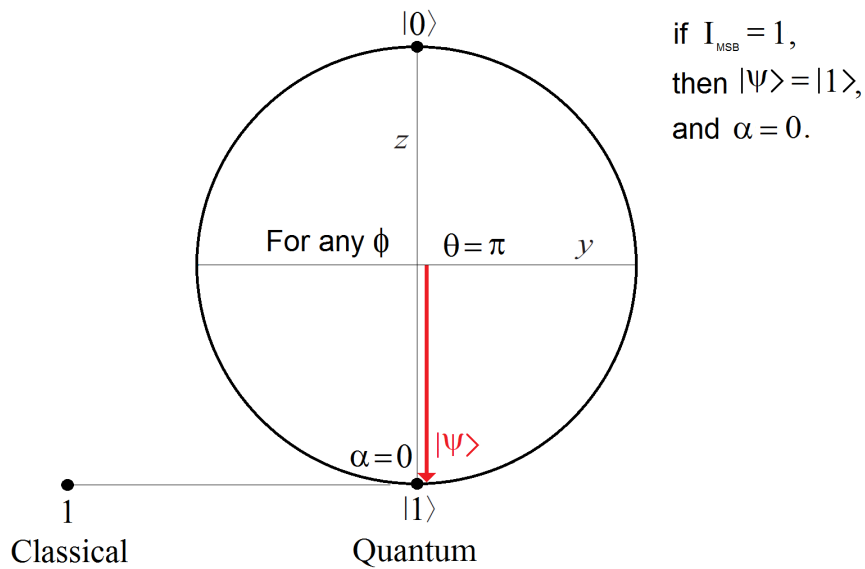

**Fig. 6** Relationship between classical 1, $\alpha$ and $I_{MSB}$.

On the other hand, in Fig.6, we can see the relationship between classical 1, α and $I_{MSB}$. This case is the opposite of the previous, with α = 0 when $I_{MSB}$ = 1 and $|\psi\rangle = |1\rangle$, with θ = π and for any ϕ too.

As we can see, the geometric relationship between Figures 5 and 6 is inverted. This happens to limiting values on the Bloch's sphere such as the CBS, that is to say, $|0\rangle$ and $|1\rangle$.

We use only the MSB of each color (see Figures 3 and 4), and we introduce the mentioned MSB to the C2QI. The output of such interface will go to the quantum algorithm, directly. See Fig. 7.

This interface is automatic and direct because we need the following correspondences, i.e.: $0 \rightarrow |0\rangle$ and $1 \rightarrow |1\rangle$, only. According to this, obviously, $\alpha = 1 - I_{MSB}$, this task is performed by a classical investor, see Fig. 7. Therefore, we obtain the rest of the wave function component as follows: $|\beta| = \sqrt{1-|\alpha|^2}$, for any ϕ (see Eq. 3 and Fig. 1 from Subsection 2.1). This latter task is performed by an actuator, which builds a wave function $\psi_{in}$ considering that the key factor of its task is the projection on the z axis (for this reason it is called actuator$_z$), i.e., α. As we can see in Figures 5 and 6, α is inverted with respect to $I_{MSB}$, i.e., if $I_{MSB}$ = 0, then α = 1 when, and $|\psi\rangle = |0\rangle$, however, if $I_{MSB}$ = 1, then α = 0 when and $|\psi\rangle = |1\rangle$. That is to say, it is only necessary to consider the z axis of Bloch's sphere when we work with CBS (see Figures 5 and 6), i.e., when we working with one qubit only, which is all that is used in this technology.

Everything mentioned here, not only facilitates the construction of future interfaces, but also makes them more simple and robust while maintaining the quality of processing within the quantum computer, for both quantum image processing [18-23] and quantum signal processing [29, 30].

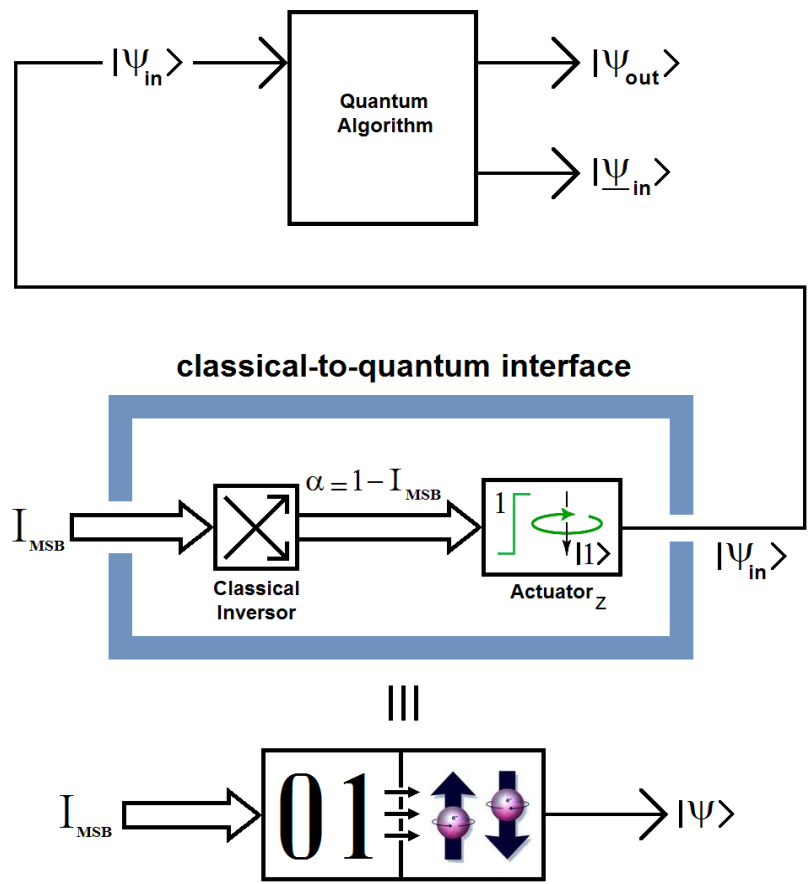

**Fig. 7** Classical-to-quantum interface.

## 4.3 Quantum Boolean image denoising

In this section, we present a method for quantum Boolean image denoising called quantum Boolean mean filter (QBMF), which works on an internal representation (inside quantum computer) of bitplane 7 (or MSB) for each color component of classical noisy image.

We present here its Boolean versions alone for simplicity in notation. Considering the above features of previous sections, the extension to the quantum Boolean version of this method is automatic. However, some preliminary considerations are necessary to understand the proceeding based on a convolutive mask.

*Convolutive mask*
In both cases (i.e., classical Boolean and quantum Boolean) we use an algorithm based on a convolutive masks with a horizontal rafter (see Fig. 8) on that $I_{MSB}$ to which we must make a denoising [31-33].

The main idea is to make an interaction between the mask and a portion of the image to be processed (with the same dimension as the mask) and that the result of said interaction to replace central pixel value of the image portion affected by the mask [14-17].

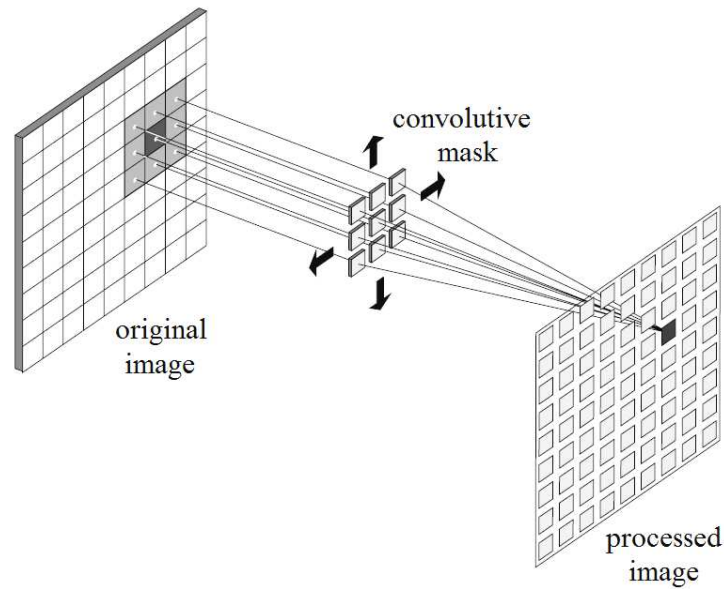

**Fig. 8** The convolution between the mask and the original image in a horizontal rafter produce the processed $I_{MSB}$.

*Quantum Boolean Mean Filter*
Based on Fig.9, we take a mask of 3×3 (often called *kernel,* which should be of any size, that is, not only 3×3, provided it has the same number of rows and columns and the dimension is an odd number) which is applied in a horizontal rafter way.

This algorithm involves three steps based on Fig.9, namely:
1. Let's calculate the number of elements in the kernel, i.e., h
2. Let's calculate the number of 1s in each kernel, i.e., n
3. Let's n and h for each pair (r,c), being r (row) and c (column) for each pixel of Imsb
4. Let's replace original Imsb(r,c) with the new result, i.e., Imsb2(r,c)

Finally, we present the classical Boolean version of mean filter in MATLAB® code instead of quantum Boolean mean filter, the reason for this is the simplicity in the notation. However, we must remember that logic is inverted inside and outside quantum computer.

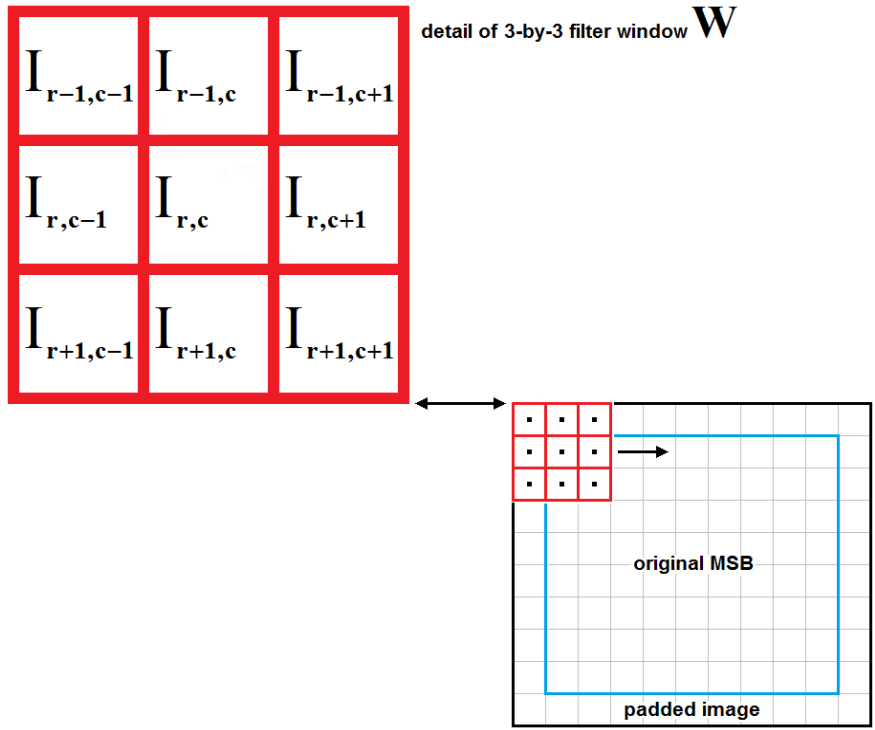

**Fig. 9** An example of 3x3 filter window for convolutive mask algorithm on a $I_{MSB}$.

```
function Imsb2 = qbmf(Imsb)

% Casting of algorithm:
% w = Mask for rafter (window or kernel)
% h = half the number of elements in the kernel
% rw = row kernel
% cw = column kernel
% n = number of 1s in each kernel
% Imsb = incoming quantum Boolean MSB for each color
% Imsb2 = outging quantum Boolean MSB for each color

Imsb2 = Imsb;
[ROW,COL] = size(Imsb);
w = input('w = ');
h = round(w*w/2);
for r = 1+floor(w/2):ROW-floor(w/2)
  for c = 1+floor(w/2):COL-floor(w/2)
    for rw = 1:w
      for cw = 1:w
        W(rw,cw) = Imsb(r-(1+floor(w/2))+rw,c-(1+floor(w/2))+cw);
      end
    end
    n = sum(sum(W));
    if(n >= h)
      Imsb2(r,c) = 1;
    else
      Imsb2(r,c) = 0;
    end
  end
end

return;
```

## 4.4 Quantum-to-classical interface (Q2CI)

In here, we recover each denoised $|\psi\rangle$ from quantum algorithm, and we introduce it to the Q2CI. The output of such interface will be the $I_{MSB}$ of each color, which will be used alongside the other bitplanes to reconstruct each color component of the denoised image, and then we turn to reconstruct the entire final denoised image. See Fig. 10.

As we can see in previous sections, there is a direct and automatic correspondence between [0, 1] and [$|0\rangle$, $|1\rangle$]. Such correspondence (and in that order) will be the classical-to-quantum interface. In the same way, but in reverse order, there is a direct and automatic correspondence between [$|0\rangle$, $|1\rangle$] and [0, 1]. In this correspondence, but in that order, we know it as a quantum-to-classical interface. Unlike [34], the measurement is not a problem, since it does not alter the outcome measure. Therefore, it is not necessary and estimator after measurement as in [34].

In Table I we can see these statements, where left column represents the state before measurement, while right column represents the state after that, for CBS and generic state (Eq.7).

TABLE I
MEASUREMENT OUTCOME WITH CBS AND GENERIC STATE.

| Before quantum measurement | After quantum measurement |
|---|---|
| $|0\rangle$ | $|0\rangle$ |
| $|1\rangle$ | $|1\rangle$ |
| $|\psi\rangle$ | $|\psi\rangle_{pm} = \dfrac{\hat{M}_m |\psi\rangle}{\sqrt{\langle\psi|\hat{M}_m^\dagger \hat{M}_m|\psi\rangle}}$ |

In Fig. 10, we can see first the quantum algorithm, whose output is directed to the interface, which begins with the measurement operator, which measure the projection on the $z$ axis (for this reason it is called measurement$_z$), i.e., $\alpha$. That is to say, it is only necessary to measure the $z$ axis as in the case of the previous interface.

The Q2CI continuous with a classical inverter, i.e., $I_{MSB} = 1 - \alpha$. This latter task is performed by a classical inversor, see Fig. 10.

As we can see in Figures 5 and 6, $\alpha$ is inverted with respect to $I_{MSB}$, i.e., if $I_{MSB} = 0$, then $\alpha = 1$ when, and $|\psi\rangle = |0\rangle$, however, if $I_{MSB} = 1$, then $\alpha = 0$ when and $|\psi\rangle = |1\rangle$. That is to say, here too, it is only necessary to consider the $z$ axis of Bloch's sphere when we work with CBS (see Figures 5 and 6), i.e., when we working with one qubit only, which is all that is used in this technology.

Table I is the cornerstone of this methodology called quantum Boolean image processing in general, and quantum Boolean image denoising in particular. It will allow us (among others):

a) to build more robust interfaces with respect to measurement noise (decoherence [35-42]),
b) to ignore the problem of quantum measurement [1, 22, 43], which was above mentioned,
c) a lower computational and memory cost [28], working only with MSB, and
d) to export this criterion beyond the quantum image processing [29, 30].

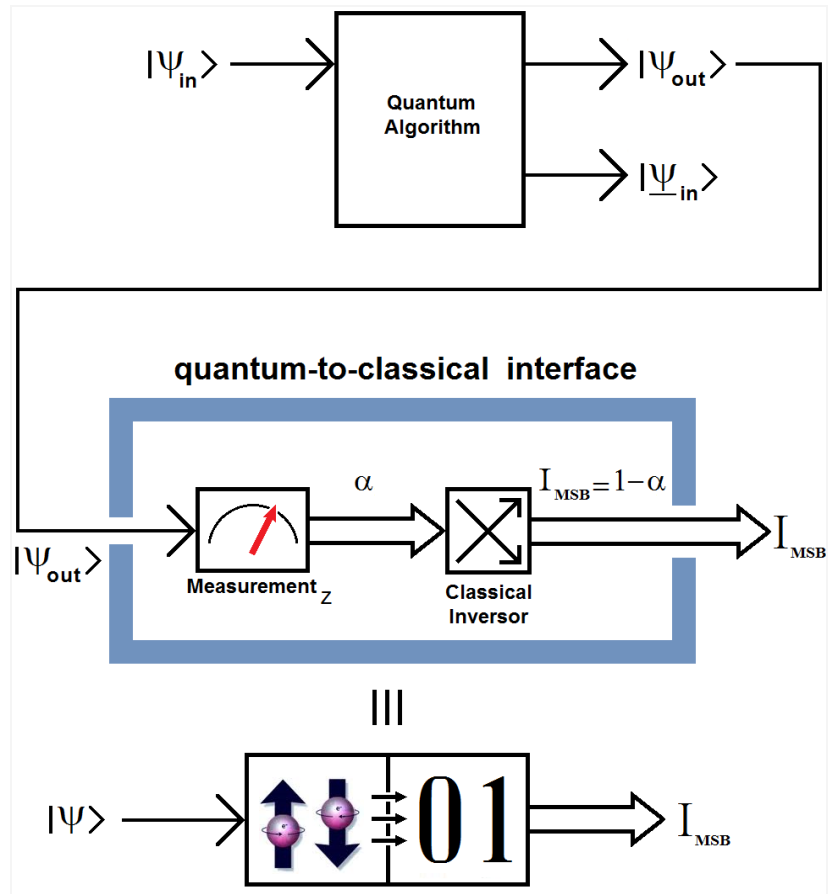

**Fig. 10** Quantum-to-classical interface.

4.5 Bit reassembling and color recomposition

We take each denoised $I_{MSB}$ and its corresponding remaining untouched bitplanes and we reassemble each color component with them, thus, and with the latter we recompose the image. This latter task is performed by a own MATLAB® function [27] called *reassembler()*, see the following code:

```
function I = reassembler(Ibpp)

% Casting of algorithm:
% bpp = bit-per-pixel
% I = Each color component of the image
% Ibpp = I in bpp bitplanes (strictly binary)
% bvpp = binary vector per pixel

[ROW,COL,bpp] = size(Ibpp);
for r = 1:ROW
  for c = 1:COL
    for b = 1:bpp
      bvpp(b) = Ibpp(r,c,b);
    end
    I(r,c) = b2d(bvpp)+1;
  end
end

return;
```

```
function p = b2d(bvpp)

% Casting of algorithm:
% bpp = bit-per-pixel
% p = pixel value
% bvpp = binary vector per pixel

bpp = length(bvpp);
p = 0;
for b = 1:bpp
   p = p + bvpp (b) * 2^(bpp-b);
end

return;
```

# 5 Metrics and Simulations

In in this section, we present a set of metrics for these experiments which are well knowledge in Digital Image Processing [14-17], and which consists in the comparison between original vs classical mean filtered and original vs quantum-Boolean mean filtering algorithms, outside and inside quantum computer, respectively.

## 5.1 Metrics

Below, we present the most conspicuous metrics used in Digital Image Processing [14-17].

*Mean Absolute Error (MAE)*
This is a conspicuous metric for these cases, which it is a quantity used to measure how close forecasts or predictions are to eventual outcomes. The mean absolute error (MAE) for gray scale images is given by

$$MAE = \frac{\sum_{r,c}|I_{original}(r,c) - I_{denoised}(r,c)|}{R \times C} \qquad (9)$$

which for two $R \times C$ (rows-by-columns) images $I_{original}$ and $I_{denoised}$, where $I_{denoised}$ means classical processed image, or quantum processed image, interchangeably.

*Mean Square Error (MSE)*
*MSE* indicates average square error of the pixels throughout the image between the original image $I_{original}$ and the classical or quantum processed image $I_{denoised}$, see Figures 11 and 12. A lower *MSE* indicates a smaller difference between both images. This means that there is a significant filter concordance. Nevertheless, it is necessary to be very careful with the edges. The formula for the *MSE* calculation for gray scale images is

$$MSE = \frac{\sum_{r,c}\left(I_{original}(r,c) - I_{denoised}(r,c)\right)^2}{R \times C} \qquad (10)$$

Here $R \times C$ pixels is the size of the images too, including original image *I*.

*Peak Signal-To-Noise Ratio (PSNR)*
PSNR is a term for the ratio between the maximum possible power of an $I_{original}$ and the power of corrupting difference that affects the fidelity of the classical or quantum processed image representation regarding original representation. Because many $I_{original}$ have a very wide dynamic range, PSNR will be expressed in terms of the logarithmic decibel scale.

We will use it as a measure of quality of coincidence between original and classical or quantum denoised versions. It is most easily defined via the mean squared error (MSE) which for two $R \times C$ (rows-by-columns) gray scale images $I_{original}$ and $I_{denoised}$, that is to say:

$$PSNR = 10\log_{10}\left(\frac{\max(I_{original})^2}{MSE}\right) = 20\log_{10}\left(\frac{\max(I_{original})}{\sqrt{MSE}}\right) \qquad (11)$$

Here, $max(I_{original})$ is the maximum pixel value of the image. When the pixels are represented using 8 bits per sample, this is 255. More generally, when samples are represented using linear pulse code modulation

(PCM) with *B* bits per sample, maximum possible value of $max(I_{original})$ is $2^B$-1. For colour images with three red-green-blue (RGB) values per pixel, the definition of PSNR is the same except the MSE is the sum over all squared value differences divided by image size and by three.

Typical values for the PSNR are between 30 and 50 dB, where higher is better.

5.2 Simulations

In Fig. 11, we can see the complete classical image denoising procedure, which will serve to any type of convolution mask filter.

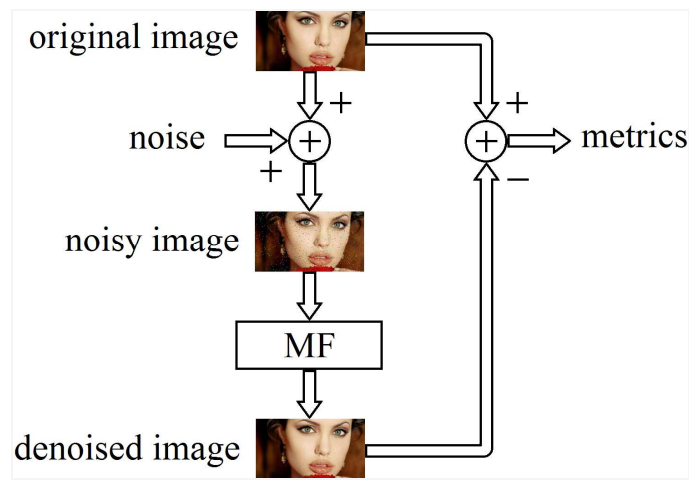

**Fig. 11** Complete classical image denoising, where, MF means mean filtering.

However, in this case we use the classical mean filter. The idea of mean filtering is simply to replace each pixel value in an image with the mean (`average') value of its neighbors, including itself. This has the effect of eliminating pixel values which are unrepresentative of their surroundings. Mean filtering is usually thought of as a convolution filter. Like other convolutions it is based around a kernel, which represents the shape and size of the neighborhood to be sampled when calculating the mean. Often a 3×3 square kernel is used, as shown in Figure 12, although larger kernels (*e.g.* 5×5 squares) can be used for more severe smoothing. (Note that a small kernel can be applied more than once in order to produce a similar but not identical effect as a single pass with a large kernel.)

| $\frac{1}{9}$ | $\frac{1}{9}$ | $\frac{1}{9}$ |
|---|---|---|
| $\frac{1}{9}$ | $\frac{1}{9}$ | $\frac{1}{9}$ |
| $\frac{1}{9}$ | $\frac{1}{9}$ | $\frac{1}{9}$ |

**Fig. 12** 3×3 averaging kernel often used in mean filtering.

For these experiments, all images are subjected to MATLAB® functions explained before, plus others built-in functions which (among other) separate the original image into its color components [27], e.g., the noise was generated using a MATLAB® R2014a (Mathworks, Natick, MA) [27] built-in function called *imnoise*. The noise type was *salt & pepper*, with a noise density of 0.05.

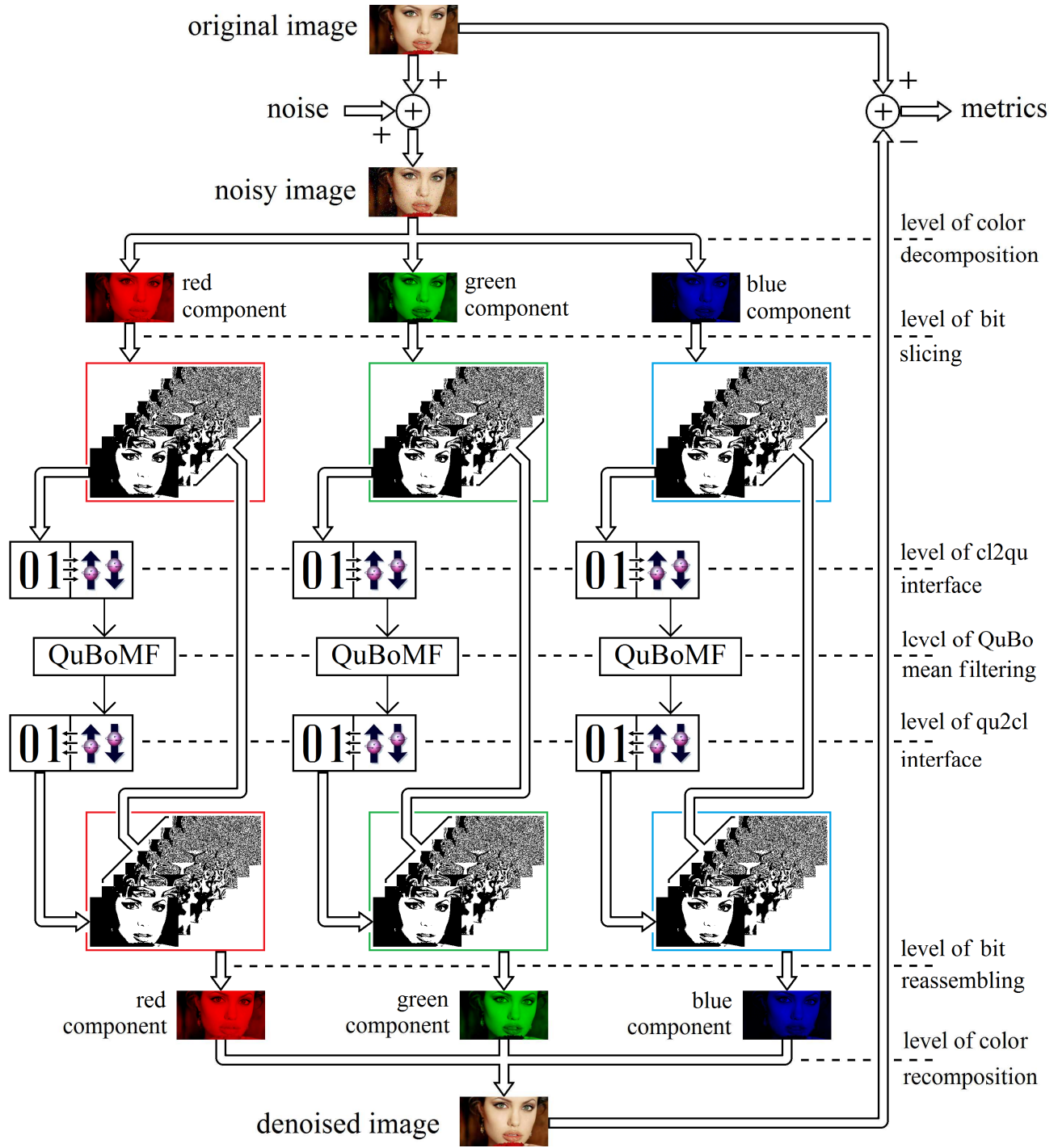

**Fig. 13** Complete quantum Boolean image denoising (for color images), where, QuBo, QuBoMF, cl2qu, and qu2cl means quantum Boolean, quantum Boolean mean filtering, classical to quantum, and quantum to classical, respectively. We must note that: 1) we work with bitplane 7 (MSB) only, while the remaining 7 directly passed to the reconstruction process, i.e., not even they are affected by interfaces, and, 2) converters have built-in interfaces, however, we do not show them to avoid complicating the figure.

In Fig. 13, we can see the complete quantum Boolean image denoising (for color images). Besides, we can appreciate the seven levels decomposition, processing and recomposition of the noisy image. In detail, we can observe level of: color decomposition, bit-slicing, classical-to-quantum interface, quantum Boolean mean filtering, quantum-to-classical interface, bit-reassembling, and finally, color recomposition.

On the other hand, first image is *Agus in Miami* (Fig. 14), which is a color Bitmap File Format (lossless) [44] of 1326-by-1326 pixels with 24 bit-per-pixel (bpp).

Fig. 14 (top-left) shows us the original image used in this experiment; noisy image (top-right); the filtered images, processed by using classical mean filter (middle-left), and quantum Boolean mean filter techniques (middle-right), respectively. Besides, Fig. 14 (down-center) shows the difference pixel-to-pixel between classical denoised vs original (noiseless) and quantum Boolean denoised vs original (noiseless), too. As we can see, there are values of pixels where the difference between two versions is remarkably sensitive.

Fig. 15 (top-left) shows us the original noiseless $I_{MSB}$ (from red color component) used in this experiment; noisy $I_{MSB}$ (top-right); the denoised $\alpha$ processed by using quantum Boolean mean filter (down-left), and the denoised $I_{MSB}$ (down-right), respectively.

In Table II, we can see MAE, MSE and PSNR results for classical and quantum Boolean among original and denoised images. The results are slightly better quantum version than the classical version.

This difference in favor of the quantum version is telling us a mismatch between the classical representation of the mean filtering and its quantum Boolean counterpart. This can only be because the noise is concentrated almost exclusively (but fully) in the MSB which is where operates the quantum Boolean version.

Second image is *Angelina* (Fig. 16), which is a color Bitmap File Format (lossless) of 1348-by-1078 pixels with 24 bit-per-pixel (bpp).

We have the same noise as in the previous case.

Fig. 16 (top-left) shows us the original image used in this experiment; noisy image (top-right); the filtered images, processed by using classical mean filter (middle-left), and quantum Boolean mean filter techniques (middle-right), respectively. Besides, Fig. 16 (down-center) shows the difference pixel-to-pixel between classical denoised vs original (noiseless) and quantum Boolean denoised vs original (noiseless), too. As we can see, there are values of pixels where the difference between two versions is remarkably sensitive here too. However, such is less than in the previous case. It has to do with a lower edges richness and texture level of *Angelina* vs *Agus in Miami*. Others important responsible factors for this difference are constituted by: a) *Agus in Miami* has higher values in its LUMA [14-17], b) *Agus in Miami* has more brightness and contrast; and, c) *Agus in Miami* is larger than *Angelina*.

This later attribute seems irrelevant to naked eye, however, it is not, since, to process more qubits, it automatically increases the detrimental intervention of bad (or poorly) modeled noise. Besides, a larger image means more openness in the time window of the process which may be more exposed to quantum decoherence [22, 35-42]. This is a topic that should be further investigated if we want to process images of very high resolution in a quantum computer.

Fig. 17 (top-left) shows us the original noiseless $I_{MSB}$ (from red color component) used in this experiment; noisy $I_{MSB}$ (top-right); the denoised $\alpha$ processed by using quantum Boolean mean filter (down-left), and the denoised $I_{MSB}$ (down-right), respectively.

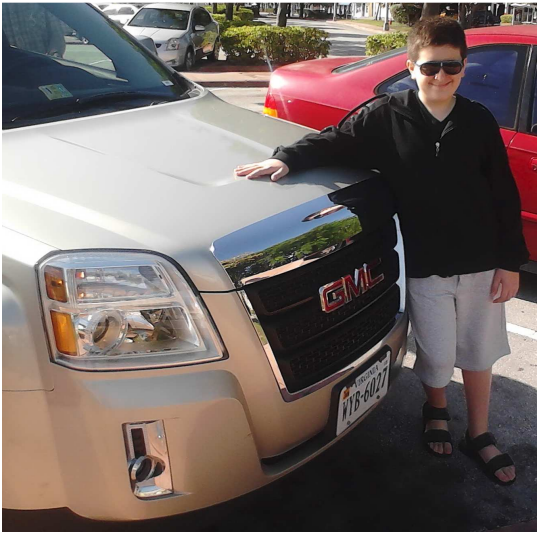
Original

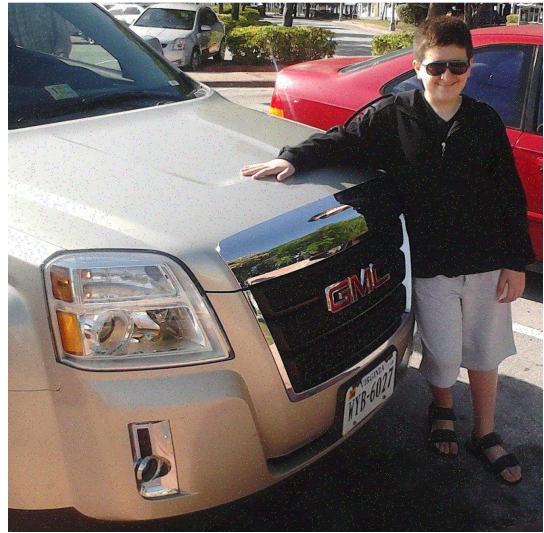
Noisy

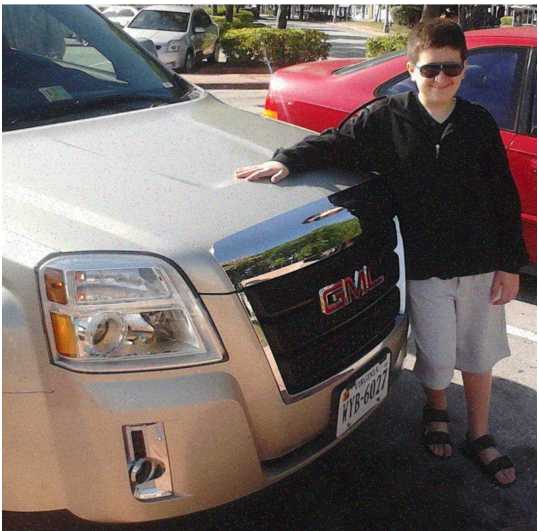
Classical denoising

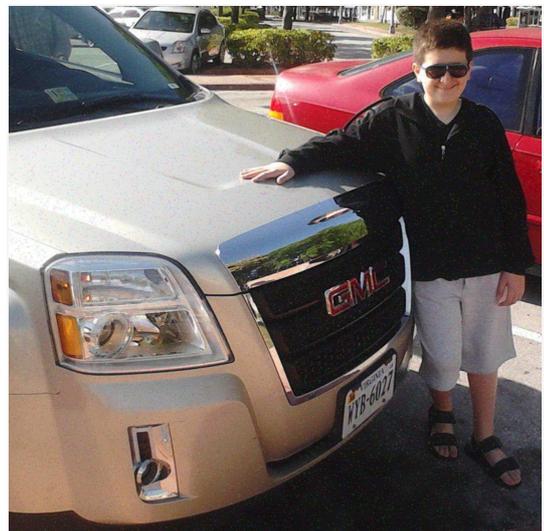
Quantum-Boolean denoising

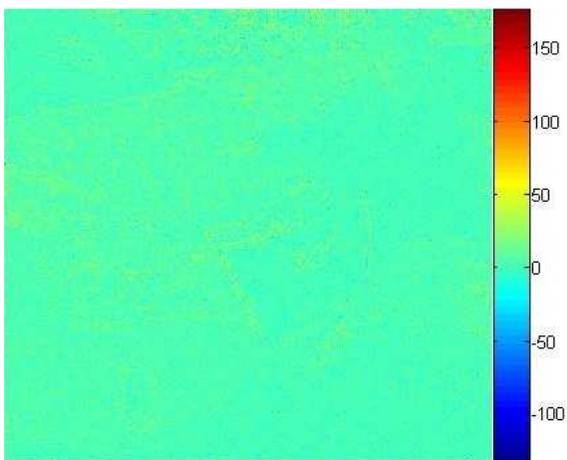
Error pixel-to-pixel of the red component for classical denoising

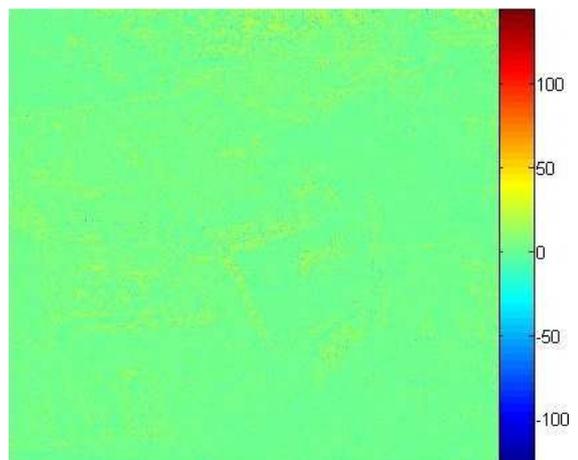
Error pixel-to-pixel of the red component for quantum-Boolean denoising

**Fig. 14** Denoising for Agus in Miami.

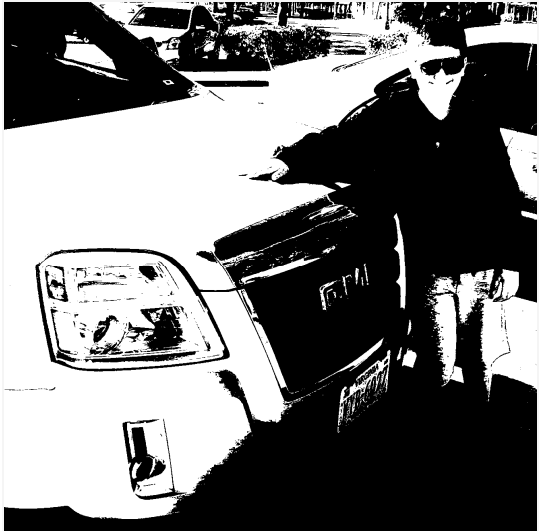 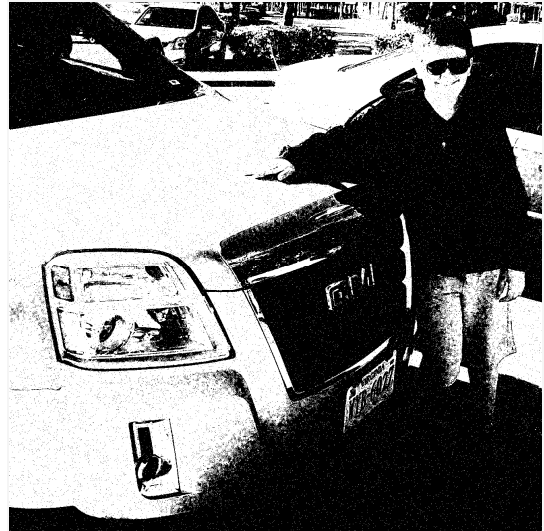

I<sub>MSB</sub> (original)          I<sub>MSB</sub> (noisy)

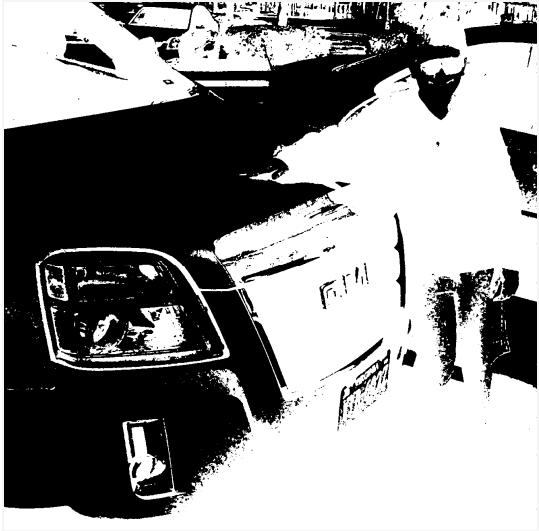 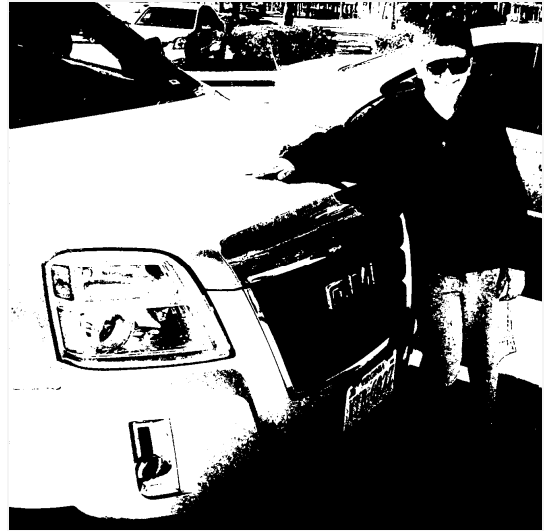

$\alpha$ (denoised)          I<sub>MSB</sub> (denoised)

**Fig. 15** Bitplane 7 (MSB) of the red component for Agus in Miami.

TABLE II
METRIC OF DENOISING FOR AGUS IN MIAMI: CLASSICAL VS. QUANTUM-BOOLEAN

| METRIC | CLASSICAL | QUANTUM-BOOLEAN |
|--------|-----------|-----------------|
| MAE    | 2.4908    | 2.0228          |
| MSE    | 20.3573   | 15.1819         |
| PSNR   | 35.0436   | 36.3175         |

In Table III, we can see MAE, MSE and PSNR results for classical and quantum Boolean among original and denoised images. Here too, the results are slightly better quantum version than the classical version.

Finally, last image is *Lena* (Fig. 18), which is a color Bitmap File Format (lossless) of 512-by-512 pixels with 24 bit-per-pixel (bpp).

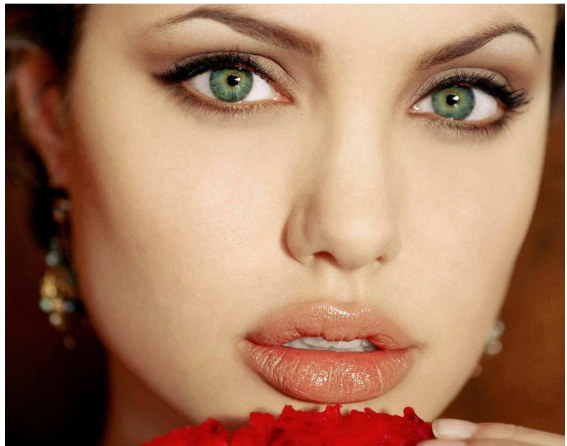
Original

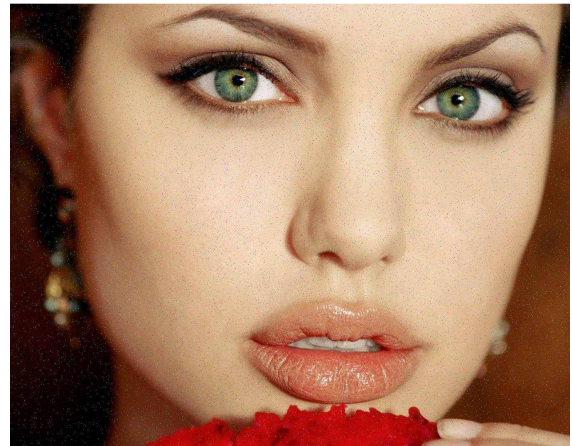
Noisy

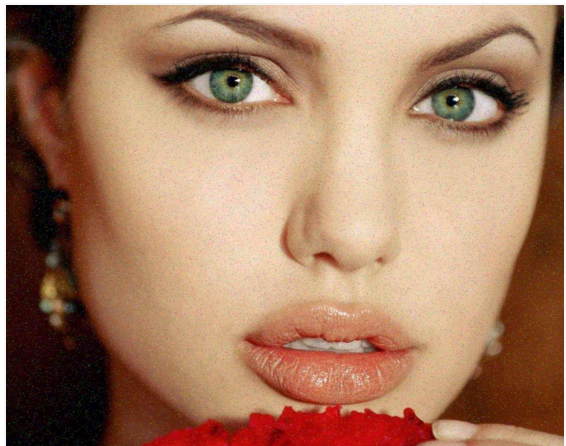
Classical denoising

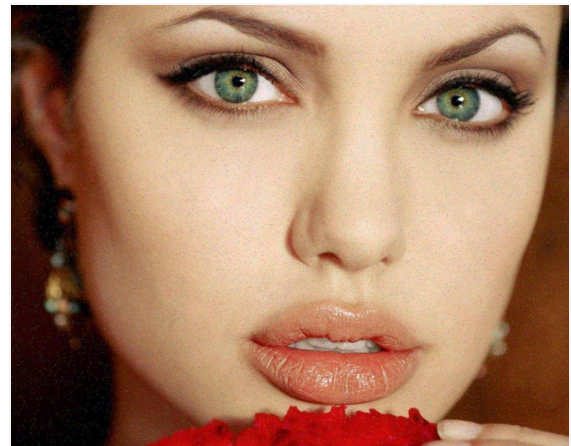
Quantum-Boolean denoising

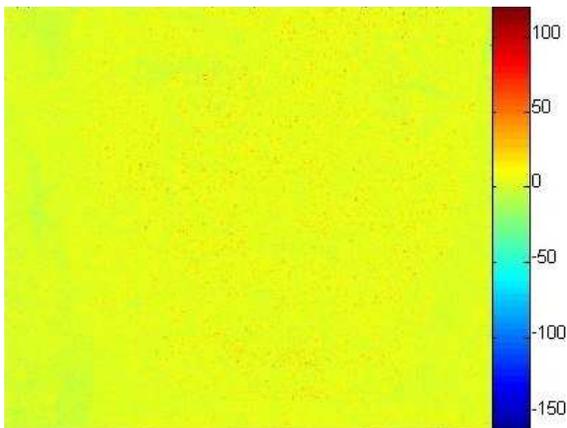
Error pixel-to-pixel of the red component
for classical denoising

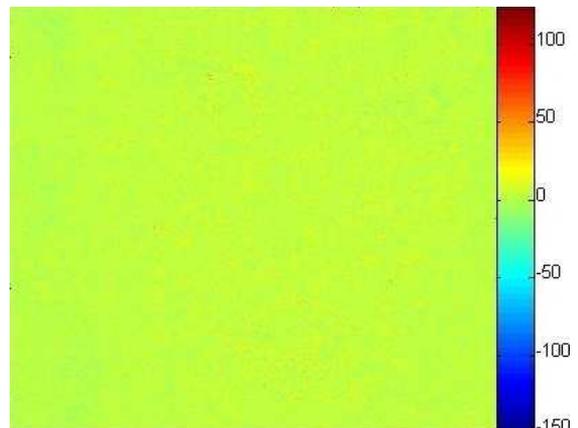
Error pixel-to-pixel of the red component
for quantum-Boolean denoising

**Fig. 16** Denoising for Angelina.

In this case, identical considerations to previous cases are used regarding to present noise in the image. Tests with other types of noise gave identical comparative results

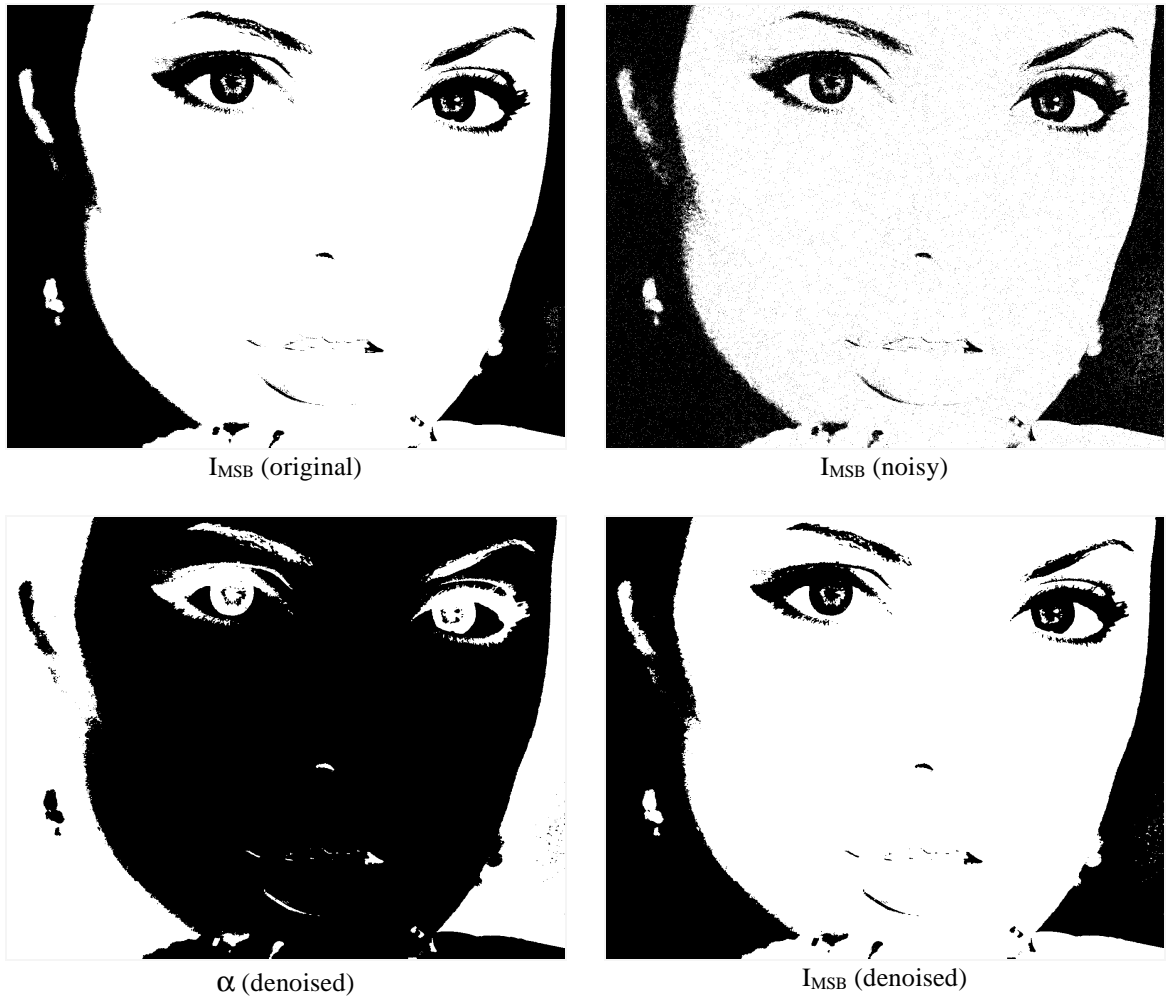

**Fig. 17** Bitplane 7 (MSB) of the red component for Angelina.

TABLE III
METRIC OF DENOISING FOR ANGELINA: CLASSICAL VS. QUANTUM-BOOLEAN

| METRIC | CLASSICAL | QUANTUM-BOOLEAN |
|---|---|---|
| MAE | 2.2284 | 1.6471 |
| MSE | 17.5597 | 10.7488 |
| PSNR | 35.6856 | 37.8172 |

Fig. 18 (top-left) shows us the original image used in this experiment; noisy image (top-right); the filtered images, processed by using classical mean filter (middle-left), and quantum Boolean mean filter techniques (middle-right), respectively. Besides, Fig. 18 (down-center) shows the difference pixel-to-pixel between classical denoised vs original (noiseless) and quantum Boolean denoised vs original (noiseless), too.

Fig. 19 (top-left) shows us the original noiseless $I_{MSB}$ (from red color component) used in this experiment; noisy $I_{MSB}$ (top-right); the denoised α processed by using quantum Boolean mean filter (down-left), and the denoised $I_{MSB}$ (down-right), respectively.

In Table IV, we can see MAE, MSE and PSNR results for classical and quantum Boolean among original and denoised images. Here too, the results are slightly better quantum version than the classical version.

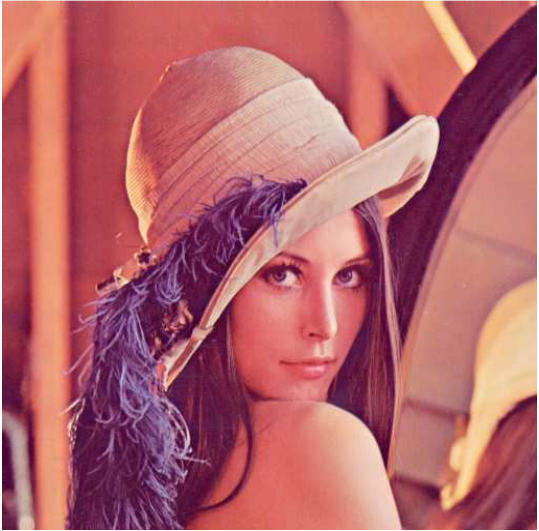
Original

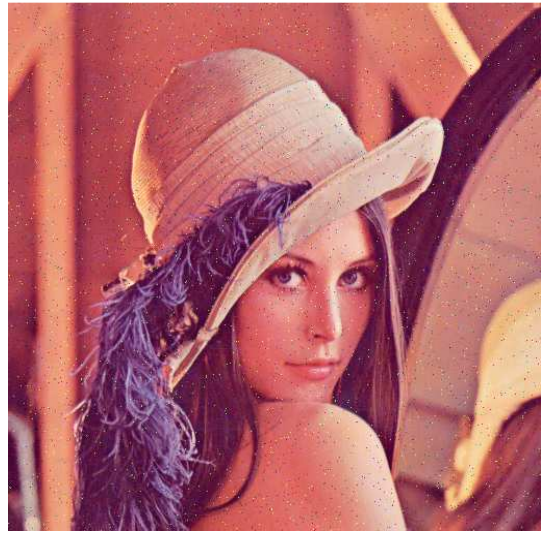
Noisy

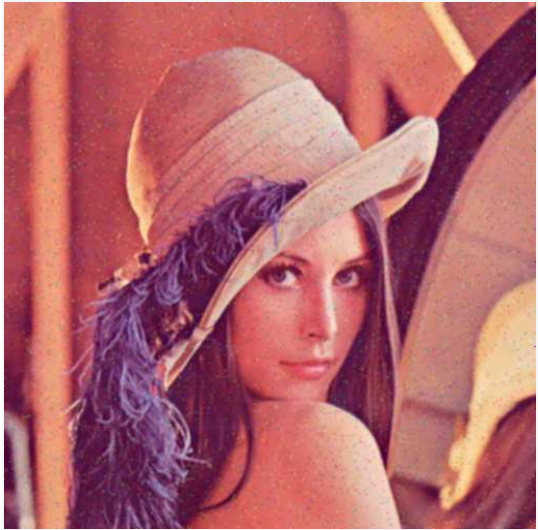
Classical denoising

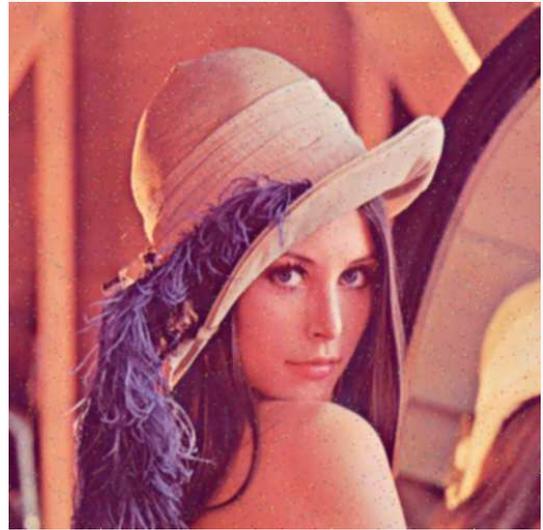
Quantum-Boolean denoising

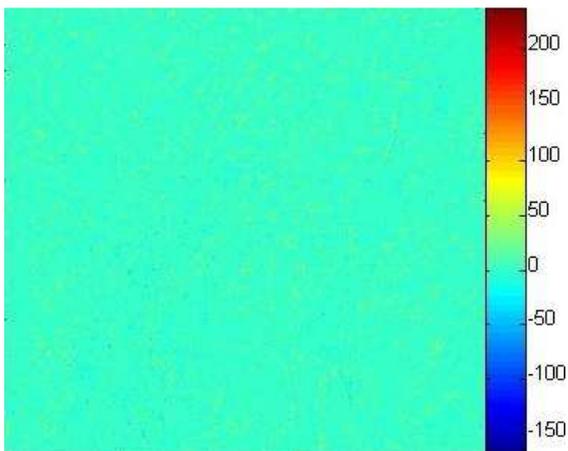
Error pixel-to-pixel of the red component
for classical denoising

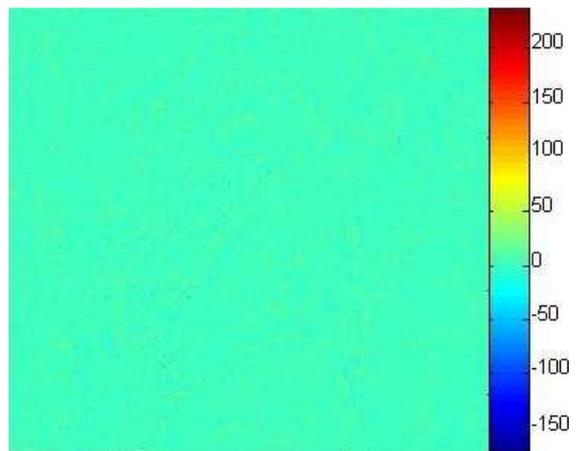
Error pixel-to-pixel of the red component
for quantum-Boolean denoising

**Fig. 18** Denoising for Lena.

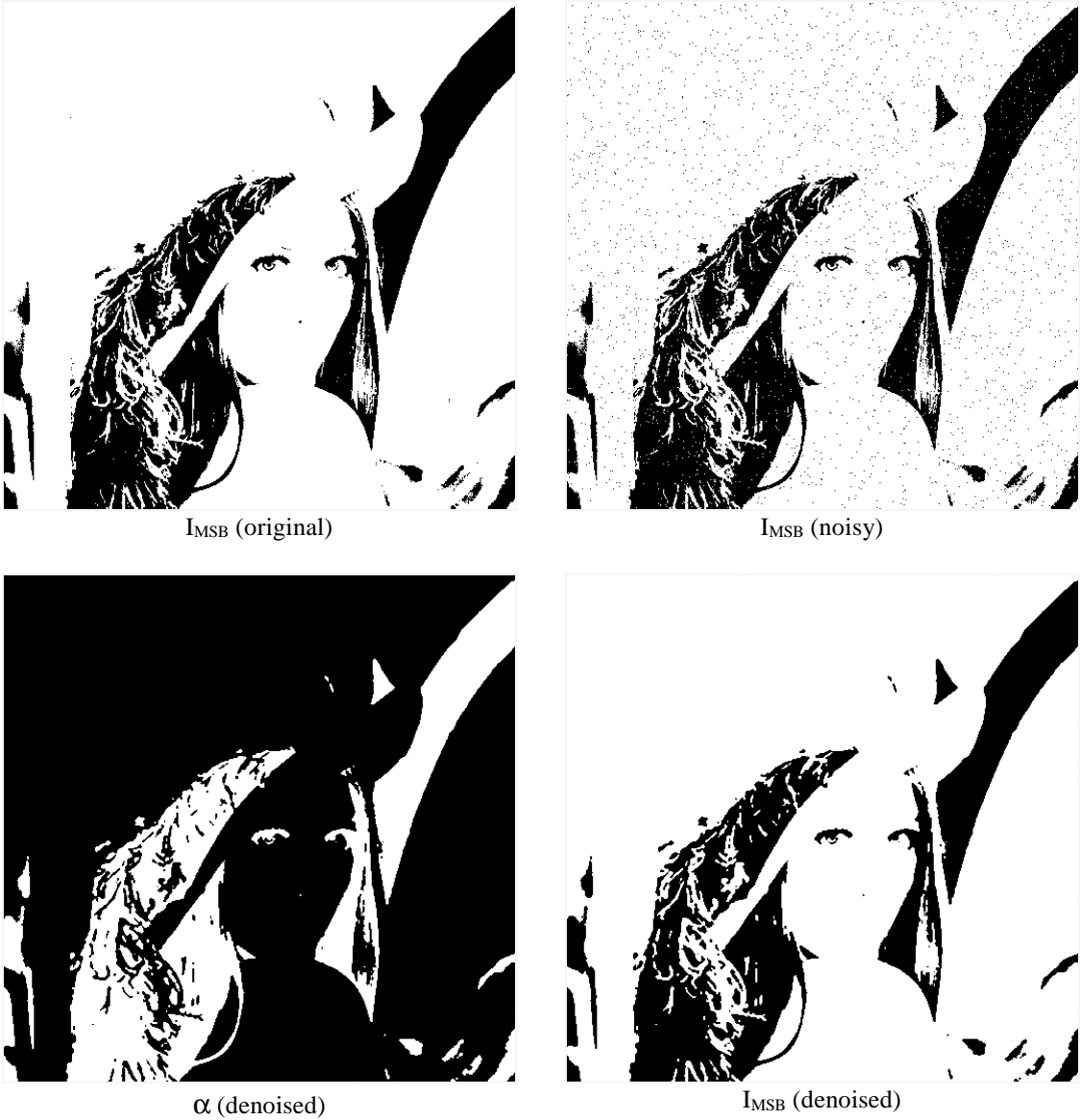

$I_{MSB}$ (original)  $I_{MSB}$ (noisy)

α (denoised)  $I_{MSB}$ (denoised)

**Fig. 19** Bitplane 7 (MSB) of the red component for Lena.

TABLE IV
METRIC OF DENOISING FOR LENA: CLASSICAL VS. QUANTUM-BOOLEAN

| METRIC | CLASSICAL | QUANTUM-BOOLEAN |
|---|---|---|
| MAE | 2.8560 | 2.4055 |
| MSE | 25.0223 | 19.6892 |
| PSNR | 34.1475 | 35.1885 |

Based on the analysis of the comparison between *Agus, Angelina* and *Lena*, we can understand why *Lena* is showing the best fit between classical and quantum Boolean version of filters. This can be seen clearly in the metrics of Table II, III and IV, where we obtain the lower difference between MAE and MSE among noiseless original and the respective denoised versions from three images so far treated. In return, this image has the highest difference value of PSNR among noiseless original and the respective denoised versions from three images so far treated. The reasons are clear: a) *Lena* has lower values in its LUMA [14-17], b) *Lena* has less brightness and contrast; and, c) *Lena* is the smallest.

# 6 Conclusions and Future Works

A quantum Boolean image denoising methodology was presented in this work. A classical Boolean version of such methodology was presented too. As we have seen, the quantum Boolean version of the filter works with computational basis states (CBS), exclusively. To achieve this, we first decompose the image into its three color components, i.e., red, green and blue. Then, we get the bitplanes for each color, e.g., 8 bits-per-pixel, i.e., 8 bitplanes-per-color. From then on, we work with the bitplane corresponding to the most significant bit (MSB) of each color, exclusive manner. After a classical-to-quantum interface (which includes a classical inverter), we have a quantum Boolean version of the image within the quantum machine. This methodology (which works with CBS, no other) allowed us to avoid the problem of quantum measurement, which alters the results of the measured except in the case of CBS. Summing-up, this methodology will enable: 1) a simpler development of logic quantum operations, where they will closer to those used in the classical logic operations, and 2) building simple and robust classical-to-quantum and quantum-to-classical interfaces. Said so far is extended to quantum algorithms outside image processing too. After filtering of the inverted version of MSB (inside quantum machine) the result passes through a quantum-classical interface (which involves another classical inverter) and then proceeds to reassemble each color component and finally the ending filtered image. Finally, this methodology minimizes the impact of decoherence [45-50], not only for quantum image denoising but also for quantum image segmentation [51].

In a special section on metrics and simulations, we use: mean absolute error (MAE), mean squared error (MSE), and peak signal-to-noise ratio (PSNR) as metrics to compare the original noiseless image vs its denoised versions, i.e., classical and quantum Boolean. The chosen denoising methods for simulations were the classic mean filter and its quantum Boolean version. The results of both simulations (outside and inside of quantum computer, respectively) show the existence of notable differences between them, which are obvious considering that each version of the algorithm is implemented in completely different spaces. However, although they were different, the quantum Boolean results were superior to those obtained by the classical technique, that is to say, all images gave more appropriate metric values. The latter is clearly seen in the computer simulations.

Clearly, the next step is the application of this technique to signal and video processing on a quantum computer [29, 30], and thus, we may exploit all its computational power and huge storage capacity. It is right to think that quantum computers will have more and more ability to absorb the computational cost of an almost unlimited number of operations in unit time in the near future. This way, quantum technology will become the main platform for multimedia real-time implementations (such as processing, storage and transmission of music and video) which combined with quantum cryptography will allow it to become the epitome of multimedia on mobile. Thus, the new quantum technology will get move completely to the current digital technology in every today imaginable and unimaginable application.

Finally, the classical technique (i.e., mean filtering) was implemented in MATLAB® R2014a (Mathworks, Natick, MA) [27] on a notebook with Intel® Core(TM) i5 CPU M 430 @ 2.27 GHz and 6 GB RAM on Microsoft® Windows 7© Home Premium 32 bits. Besides, a simulated version of quantum implementations were done on a GPU cluster, NVIDIA® Tesla© 2050 GPU [52] with a peak performance of approximately 500 GFLOPS, with an achieved performance of approximately 250 GFLOPS in OpenCL. The GPU needed approximately 2.5 GB of bandwidth with InfiniBand connectivity at quad data rate (QDR) QLogic® [53] or 40 Gb speeds.

**Acknowledgments** M. Mastriani thanks Prof. Dr. Salvador E. Venegas-Andraca, Assistant Professor of Mathematics and Computer Science from Monterrey Institute of Technology-State of Mexico Campus for being the creator of this discipline, by share it -so selflessly- with all humanity, and for his tremendous help and support. Besides, I wish to thank all the technical staff of the various laboratories of the National Commission of Atomic Energy for the help they gave me in the preparation of experiments. It is impossible to name them all here, simply, thank you all for all.

**References**

[1] Nielsen M. A. and Chuang I. L.: Quantum Computation and Quantum Information, Cambridge University Press, Cambridge, England, 2004.
[2] P. A. Benioff. Quantum mechanical Hamiltonian models of Turing machines. Journal of Statistical Physics, 29(3):515–546, 1982.
[3] R. Feynman. Simulating physics with computers. International Journal of Theoretical Physics, 21(6/7):467–488, 1982.
[4] R. Feynman. Quantum mechanical computers. Optics News, 11:11–20, 1985.
[5] D. Deutsch. Quantum theory, the Church-Turing principle, and the universal quantum Turing machine. In Proceedings of the Royal Society of London, volume A400, pages 97–117, 1985.
[6] D. Deutsch and R. Jozsa. Rapid solution of problems by quantum computation. In Proceedings of the Royal Society of London, volume A439, pages 553–558, 1992.
[7] D. Simon. On the power of quantum computation. SIAM Journal on Computing, 26(5):1474–1483, 1997.
[8] E. Bernstein and U. Vazirani. Quantum complexity theory. SIAM Journal on Computing,26(5):1411–1473, 1997.
[9] P. W. Shor. Polynomial-time algorithms for prime factorization and discrete logarithms on a quantum computer. SIAM Journal on Computing, 26(5):1484–1509, 1997. quant-ph/9508027.
[10] R. Rivest, A. Shamir, and L. Adleman. A method for obtaining digital signatures and public key cryptosystems. Communications of the ACM, 21:120–126, 1978.
[11] J. van Leeuwen, editor. Handbook of Theoretical Computer Science. Volume A: Algorithms and Complexity. MIT Press, Cambridge, MA, 1990, pages 717–755.
[12] C. H. Bennett and G. Brassard. Quantum cryptography: Public key distribution and coin tossing. In Proceedings of the IEEE International Conference on Computers, Systems and Signal Processing, pp.175–179, 1984.
[13] Le P.Q., Dong F., Hirota K.: A flexible representation of quantum images for polynomial preparation, image compression, and processing operations. Quantum Inf Process (2011) 10:63–84. doi:10.1007/s11128-010-0177-y
[14] Jain A. K.: Fundamentals of Digital Image Processing, Englewood Cliffs, New Jersey, 1989.
[15] Gonzalez R. C. and Woods R. E.: Digital Image Processing, 2nd Edition, Prentice- Hall, Jan. 2002.
[16] Gonzalez R. C., Woods R. E. and Eddins S. L.: Digital Image Processing using Matlab. Upper Saddle River, NJ: Pearson Prentice Hall, 2004.
[17] Schalkoff R. J.: Digital Image Processing and Computer Vision, Wiley, 1989, New York.
[18] Venegas-Andraca, S.E., Ball, J.L.: Storing Images in engtangled quantum systems. arXiv:quantph/ 0402085 (2003)
[19] Venegas-Andraca, S.E., Bose, S.: Storing, processing and retrieving an image using quantum mechanics. Proc. SPIE Conf. Quantum Inf. Comput. vol. 5105, 137–147 (2003). doi:10.1117/12.485960
[20] Venegas-Andraca S. E., Bose S.: Quantum computation and image processing: New trends in artificial intelligence. Proceedings of the International Conference on Artificial Intelligence IJCAI-03, pp. 1563-1564, 2003.
[21] Venegas-Andraca S. E., Ball J. L.: Processing images in entangled quantum systems. Springer-Verlag: Quantum Information Processing, Vol.9, No.1, pp.1-11, 2010.
[22] Venegas-Andraca S. E.: DPhil thesis: Discrete Quantum Walks and Quantum Image Processing. Centre for Quantum Computation, University of Oxford, 2006.
[23] Latorre, J.I.: Image compression and entanglement. arXiv:quant-ph/0510031 (2005)
[24] Kaye, P., Laflamme, R., Mosca, M.: An Introduction to Quantum Computing. Oxford University Press, Oxford (2004)
[25] Stolze, J., Suter, D.: Quantum Computing: A Short Course from Theory to Experiment. WILEY-VCH Verlag GmbH & Co. KGaA, Weinheim (2007)
[26] Busemeyer, J.R., Wang, Z., Townsend, J.T.: Quantum dynamics of human decision-making. Journal of Mathematical Psychology 50 (2006) 220–241. doi:10.1016/j.jmp.2006.01.003
[27] MATLAB® R2014a (Mathworks, Natick, MA). http://www.mathworks.com/
[28] M. Mastriani, "Systholic Boolean Orthonormalizer Network in Wavelet Domain for Microarray Denoising," International Journal of Signal Processing, Volume 2, Number 4, pp.273-284, 2006. [Online]. Available: https://www.waset.org/author/mario-mastriani

[29] Eldar, Y.C.: Quantum Signal Processing, Doctoral Thesis, Massachusetts Institute of Technology, December 2001.
[30] Eldar, Y.C., Oppenheim, A.V.: Quantum Signal Processing, Signal Processing Mag., vol. 19, pp. 12-32, Nov. 2002.
[31] Mastriani, M., Giraldez, A.: Enhanced Directional Smoothing Algorithm for Edge-Preserving Smoothing of Synthetic-Aperture Radar Images. Journal of Measurement Science Review, Volume 4, Section 3, pp.1-11, 2004.
[32] Mastriani, M., Giraldez, A.: Smoothing of coefficients in wavelet domain for speckle reduction in Synthetic Aperture Radar images. ICGST International Journal on Graphics, Vision and Image Processing (GVIP), Volume 6, pp. 1-8, 2005.
[33] Mastriani, M., Giraldez, A.: Despeckling of SAR images in wavelet domain. GIS Development Magazine, Sept. 2005, Vol. 9, Issue 9, pp.38-40.
[34] Mastriani, M.: Optimal Estimation of States in Quantum Image Processing. (2014) ArXiv:1406.5121 [quant-ph]
[35] Zurek, W.H.: Decoherence and the Transition from Quantum to Classical–Revisited, (2003) arXiv:quant-ph/0306072v1
[36] Alagic, G., Russell, A.: Decoherence in quantum walks on the hypercube. quant-ph/0501169 (2005)
[37] Dass, T.: Measurements and Decoherence. arXiv:quant-ph/0505070v1 (2005)
[38] Kendon, V., Tregenna, B.: Decoherence in a quantum walk on the line. Proceedings of QCMC 2002, (2002)
[39] Kendon, V., Tregenna, B.: Decoherence can be useful in quantum walks. Phys. Rev. A, 67:042315, (2003)
[40] Kendon, V., Tregenna, B.: Decoherence in discrete quantum walks. Selected Lectures from DICE 2002. Lecture Notes in Physics, 633:253–267, (2003)
[41] Romanelli, A., Siri, R., Abal, G., Auyuanet, A., Donangelo, R.: Decoherence in the quantum walk on the line. Phys. A, 347c:137–152, (2005)
[42] DiVincenzo, D. P.: The Physical Implementation of Quantum Computation. arXiv:quant-ph/0002077v3 (2008)
[43] DiVincenzo, D. P.: Quantum Computation. Science, New Series, Vol. 270, No. 5234 (Oct. 13, 1995), pp. 255-261.
[44] Miano, J.: Compressed Image File Formats: JPEG, PNG, GIF, XBM, BMP; Ed. Addison-Wesley, N.Y. (1999)
[45] Wheeler, N.: Problems at the Quantum/Classical Interface. (2001) http://ebookily.org/pdf/problems-at-the-quantum-classical-interface-174658500.html
[46] Baylis, W.E.: Quantum/Classical Interface: A Geometric Approach from the Classical Side. Computational Noncommutative Algebra and Applications. NATO Science Series II: Mathematics, Physics and Chemistry Volume 136, pp.127-154, (2004)
[47] Baylis, W.E., Cabrera, R., Keselica, D.: Quantum/Classical Interface: Fermion Spin. (2007) arXiv:0710.3144v2
[48] Svozil, K.: Quantum Interfaces. CDMTCS Research Report Series, Technische Universitat Wien, Austria, CDMTCS-136, May 2000.
[49] Landsman, N.P.: Between classical and quantum, (2005) arXiv:quant-ph/0506082v2
[50] Zhou, X., Bocko, M.F., Feldman, M.J.: Isolation Structures for the Solid-State Quantum-to-Classical Interface. Presented at International Conference on Quantum Information, March 2001, Rochester, N.Y.
[51] Mastriani, M.: Quantum Edge Detection for Image Segmentation in Optical Environments. International Journal of Optics, Hindawi (in review process). (2014)
[52] NVIDIA® Tesla© 2050 GPU. http://www.nvidia.com/
[53] QLogic® QDR Infiniband. http://www.qlogic.com/